\documentclass[aps,prb,twocolumn,showpacs,superscriptaddress]{revtex4-1}

\usepackage{amsmath}
\usepackage{graphicx}
\usepackage{hyperref}
\usepackage{color}
\usepackage{algorithm}
\usepackage{algorithmic}
\usepackage{amsfonts}
\usepackage{amssymb}

\begin{document}

\title{The Continuous-Pole-Expansion method to obtain spectra of electronic lattice models}
\author{Peter Staar}
\affiliation{Institute for Theoretical Physics, ETH Zurich, 8093 Zurich, Switzerland}
\author{Bart Ydens}
\affiliation{Laboratory of Solid-State Physics and Magnetism, KU Leuven, 3001 Leuven, Belgium}
\author{Anton Kozhevnikov}
\affiliation{Institute for Theoretical Physics, ETH Zurich, 8093 Zurich, Switzerland}
\author{Jean-Pierre Locquet}
\affiliation{Laboratory of Solid-State Physics and Magnetism, KU Leuven, 3001 Leuven, Belgium}
\author{Thomas Schulthess}
\affiliation{Institute for Theoretical Physics, ETH Zurich, 8093 Zurich, Switzerland}
\affiliation{Swiss National Supercomputing Center, ETH Zurich, 6900 Lugano, Switzerland} 
\date{\today }

\begin{abstract}
We present a new algorithm to analytically continue the self-energy of quantum many-body systems from Matsubara frequencies to the real axis. The method allows straightforward, unambiguous computation of electronic spectra for lattice models of strongly correlated systems from self-energy data that has been collected with state-of-the are continuous time solvers within dynamical mean field simulations. Using well-known analytical properties of the self-energy, the analytic continuation is cast into a constrained minimization problem that can be formulated as a quadratic programmable optimization with linear constraints. The algorithm is validated against exactly solvable finite size problems, showing that all features of the spectral function near the Femi level are very well reproduced and coarse features are reproduced for all energies. The method is applied to two well known lattice problems, the two-dimensional Hubbard model at half filling where the momentum dependence of the gap formation is studied, as well as a multi-band model of NiO, for which the spectral function can be directly compared to experiment. Agreement with results published results is very good.
\end{abstract}


\maketitle

\section{Introduction}

Studies of electronic lattice models with intermediate to strong correlations have traditionally been very important in condensed matter physics. Most relevant models are not tractable with controlled analytic approximations in the parameter regions of interest, thus requiring numerical simulations for their solution. Currently, dynamical mean field theory (DMFT)
\cite{Georges1996RMP} is the method of choice in many investigations of this type of problems. For practical reasons DMFT approaches these models in imaginary rather than real time. Consequently, physical observables, such as spectral functions, can only be accessed indirectly via analytic continuation from the imaginary to the real axis.

Two methods are commonly used for analytic continuation to the real axis. In the Pade-approximation \cite{Vidberg1977JLTP,Anisimov1997JOP} a fractional polynomial is fit to the data that has been computed on the Matsubara frequencies on the imaginary axis, and the polynomial is evaluated on the real axis. The procedure is rather general but for most physical cases requires fractional polynomials that violate known analytic properties of the fitted functions in the complex plane. The second and much more successful method is the Maximum Entropy Method (MEM)\cite{Gubernatis1991PRB,Jarrell1996PR,Silver1990PRB}. This method analytically continues the imaginary time Greens function to the real frequency axis in order to obtain the spectrum $\mathcal{A}$. It is based on the relationship

\begin{align}\label{MEM_rel}
G(\tau) &= \frac{-1}{\pi} \int d\omega \underbrace{\frac{e^{-\tau\,\omega}}{1+e^{-\beta\,\omega}}}_{=K(\omega, \tau)} \mathcal{A}(\omega).
\end{align}

A straightforward (numerical) inversion of Eq~(\ref{MEM_rel}) is impossible, since the spectrum at large frequencies ($\omega \gg 1$) has only an exponentially small contribution to the imaginary time Greens function. Thus, at finite numerical precision, there are many different spectral functions that satisfy Eq~(\ref{MEM_rel}). The central idea in MEMs is to search for a spectral function that satisfies Eq~(\ref{MEM_rel}) and maximizes the information entropy $\mathcal{S}$~\cite{Shannon1948}, relative to a positive definite function $m(\omega)$ which has the correct high-frequency behavior

\begin{align}\label{MEM_entropy}
\mathcal{S} &= -\int d\omega \mathcal{A}(\omega) - m(\omega) - \mathcal{A}(\omega) \, \log(\mathcal{A}(\omega)/m(\omega)).
\end{align}

\noindent
Here the $m(\omega)$ function serves as the default model. In absence of the constraint in Eq~(\ref{MEM_rel}), maximizing the entropy will result in a spectrum $\mathcal{A}(\omega)$ equal to $m(\omega)$. For unknown systems, finding a good default model is often not straightforward, and results from MEM-based analytic continuation do not seem transparent.   

With the introduction of continuous time Monte Carlo solvers~\cite{Rubtsov2005PRB,Werner2006PRL,Werner2006PRB,Gull2008EPL,Gull2011RMP}, it has become possible to directly measure the self-energy on the Matsubara axis with unprecedented speed~\cite{Gull2011PRB} and accuracy~\cite{Staar2012JPCS}. This motivated us to investigate the possibility of an analytic continuation of the self-energy directly in frequency space with the relationship

\begin{align}\label{CPE_rel}
\Sigma(z) &=  \Sigma_0 + \frac{1}{2\pi} \int_{-\infty}^{\infty} d\omega \underbrace{\frac{1}{\omega-z}}_{=T(\omega, z)} \, \mbox{Im} \big[ \Sigma(\omega) \big].
\end{align}

Analytic continuation in frequency is advantageous, since the ill-defined high-frequency behavior caused by the exponential decay of the kernel in Eq.~(\ref{MEM_rel}) can be avoided. This is easily demonstrated. On the Matsubara poles the high-frequency part of the self-energy behaves like

\begin{align}\label{CPE_rel2}
\Sigma(\varpi \gg 1) \approx \Sigma_0 - \imath \, \frac{\Sigma_1}{\varpi} + \cdots, \quad \Sigma_0, \Sigma_1 \in R
\end{align}

\noindent
Equating this to a high-frequency expansion of the right hand side of Eq.~(\ref{CPE_rel}) gives

\begin{align}
\Sigma_1 =  \int_{-\infty}^{\infty} d\omega \, \mbox{Im} \big[ \Sigma(\omega) \big], \qquad \mbox{Im} \big[ \Sigma(\omega) \big]\leq0.
\end{align}

\noindent
The imaginary part of the self-energy has a finite $L_1$-norm, since from standard field theory~\cite{AGD} it follows that the imaginary part of the self-energy is strictly negative on the real axis, and thus must decay on the real axis for large frequencies. The exponential decay of the transfer function $K(\omega, \tau)$ in Eq.~(\ref{MEM_rel}) for large frequencies is replaced by a polynomial decay of the new transfer function $T(\omega, \varpi)$ in Eq.~(\ref{CPE_rel}), simply by keeping the analytic continuation completely in the frequency domain. 

Despite the improvement to the ill-conditioned high-frequency problem on the real axis, inverting Eq.~(\ref{CPE_rel}) remains non-trivial for a number of practical reasons. Firstly, straight numerical inversion of the transfer-matrix $T(\omega_i, \varpi_j)$ is unstable. Secondly, in the typical case when the self-energy is obtained from a Monte-Carlo simulation, the algorithm has to be robust against statistical noise in the data \--- it should not extract information for the self-energy on the real axis from simple statistical noise. Finally, the constraint that the imaginary part of the self-energy has to be negative is difficult to enforce while solving a linear system.

For all these reasons we chose in the present paper to convert the inversion of Eq.~(\ref{CPE_rel}) into a minimization problem. The desired self-energy is described by a functional form $f$ that is strictly positive on the real axis, and the minimization function $\Lambda$ is constructed from the L2-norm of the difference to the Monte Carlo simulation data on the Matsubara frequencies. The solution that minimizes $\Lambda$ corresponds to the solution of Eq.~(\ref{CPE_rel}). 

Besides intrinsic robustness, the minimization approach has the advantage that strong constraints can easily be imposed on the targeted solution. The goal of this paper is to investigate the parametrization of the function $f$ and presenting a viable implementation of the constrained minimization problem.

A significant portion of the present work has been dedicated to the validation of the CPE algorithm. First we compare against spectra obtained via exact diagonalization (ED) solutions for isolated cluster models. With ED methods one computes the exact self-energy function on the imaginary and real axis for small problems, making a straightforward validation possible. Next we apply the CPE to self-energy data obtained for the single-band Hubbard model~\cite{Hubbard1963I} in two dimensions (2D) at half filling, where the self-energy data is computed with the Dynamical Cluster Approximation (DCA) \cite{Hettler1998PRB,Hettler2000PRB,Jarrell2001PRB,Maier2005RMP} on a 32-site cluster. We validate against a formula commonly used in the literature to probe the spectral density at the Fermi-energy. We illustrate how the CPE can be used to inspect the momentum-dependence of the spectrum by investigating the momentum dependent gap formation. This topic has been recently investigated on small clusters~\cite{Werner2009PRB,Gull2009PRB,Gull2010PRB}, but without any conclusions on the spectral functions. We will show that the CPE arrives at the same conclusions as in the literature and investigate the spectral functions more closely. Lastly, we apply the CPE to a multi-band model of NiO, a well studied material. We demonstrate how the orbital dependent spectrum can be computed with the CPE, and validate the results against experimental data available from the literature.  As there are no exact results for the impurity problem, experimental XPS, XES and BIS spectra are the next best option to validate the CPE algorithm on materials specific models. We find a remarkably good agreement between theory and experiment.

The paper is structured as follows: In section~\ref{analytic_prop} we review some important analytical properties of Matsubara Greens function and self-energy. In section~\ref{CPE} we introduce the CPE for the self-energy and treat thoroughly the numerical implementation. In section~\ref{EDresults}, we compare the CPE with the exact results obtained with the ED solution of an isolated cluster. In section ~\ref{results}, we apply CPE to some physically relevant problems, and compare the results of the CPE with the literature. Section~\ref{concsec} contains the conclusions.

\section{Analytical properties of Fermionic Green's function and self-energy}\label{analytic_prop}

In order to motivate and later derive the Continuous Pole Expansion (CPE) algorithm for the self-energy, we briefly review the analytical properties of the single particle Green's function as well as the self-energy for Fermionic systems. Following Abrikosov, Gorkov and Dzyaloshinski\cite{AGD}, the single particle propagator  $G$ is defined as

\begin{align}\label{def_G}
G(\vec{k},\tau) &= \langle T_\tau [ c_{\vec{k}}^\dagger(\tau) c_{\vec{k}}(0) ] \rangle,
\end{align}

\noindent
where the imaginary time $\tau \in [-\beta, \beta]$. Due to the time ordering operator $T_\tau$ and Fermionic commutation relations among the field operators $c^\dagger(\tau)$ and $c(\tau)$, where for simplicity we omit the momentum vector $\vec{k}$, the Green's function values for $\tau<0$ and $\tau>0$ are related by $G(\tau+\beta) = -G(\tau)$. Consequently, the Fourier transform of the Fermionic Greens-function is only non-zero on the Matsubara frequencies $\varpi_m=\: \pi/\beta\,(2\,m+1)$ with $\: m\:\in\: \mathbb{Z}$ and we have that

\begin{align}
G(\varpi_m) &= \int_0^\beta d\tau \: e^{i\:\varpi_m\:\tau}\:G(\tau).
\end{align}
  
\noindent
Since there exists a unique, analytical function that coincides with the infinite sequence  $\{\imath\,\varpi_m, G(\varpi_m)\}$ in the complex plane
\footnote{We refer here to the interior uniqueness properties of single valued complex functions from http://www.encyclopediaofmath.org: \textit{Let $\mathcal{D}$ be a domain in the complex plane $\mathbb{C}$. The classical interior uniqueness theorem for holomorphic, i.e. single-valued analytic, functions on $\mathcal{D}$ states that if two holomorphic functions $f(z)$ and $g(z)$ in $\mathcal{D}$ coincide on some set $\mathcal{E} \subset \mathcal{D}$ containing at least one limit point in $\mathcal{D}$, then  $f(z)=g(z)$ everywhere in $\mathcal{D}$.}},
we define the Greens-function $G(z)$ on the entire complex plane as the unique analytical continuation of this infinite sequence. Furthermore, a straightforward expansion of Eq.~(\ref{def_G}) in terms of the eigen-energies and eigen-basis of the system reveal that there exists a positive, integrable, real function $\rho(\omega)$ such that the Greens-function on the real axis can be obtained as, 

\begin{align}\label{AGD_rel}
G(\varpi_m) &= \int_{-\infty}^{+\infty} d\omega' \: \frac{\rho(\omega')}{\omega'-\varpi_m}.
\end{align}

\noindent
Due to uniques we can generalize Eq.~(\ref{AGD_rel}) to anywhere in the (upper) complex plane and obtain 

\begin{align}\label{AGD_gen}
G(z) &= \int_{-\infty}^{+\infty} d\omega' \: \frac{\rho(\omega')}{\omega'-z}.
\end{align}

\noindent
Since $\rho$ is a real and positive function, we can deduce from Eq.(\ref{AGD_gen}) that the following analytical properties hold for the Greens-function,

\begin{align}\label{prop_G}
\mbox{Im}\big[G(\omega + i\varpi)\big] &< 0 \: \mbox{if} \: \varpi > 0. \\
G^I(i\varpi) &= \overline{G^I(-i\varpi)} \nonumber
\end{align}

\noindent
It follows, that the single particle propagator can only have complex zeros on the real axis.

The self-energy $\Sigma$ is related to the Greens-function via the Dyson equation,

\begin{align}\label{def_Sigma}
\Sigma(z) &= G^{-1}_0(z)  - G^{-1}(z). 
\end{align}

\noindent
Since $G(z)$ and $G_0(z)$ have no zero's in the upper complex plane, it follows that their inversion can not introduce poles in the upper plane of the self-energy. Hence, the self-energy is also analytic in the upper complex plane, with the possible exception on the real axis. The absence of poles in the upper complex plane permits use of the residue theorem

\begin{align}\label{prop_S1}
\Sigma(z) =  \frac{1}{2\pi \imath}\lim_{\delta\rightarrow0^+} \int_{-\infty}^{\infty} d\omega \frac{\Sigma(\omega+\imath\,\delta)}{\omega-z}.
\end{align}

\noindent
Due to causality the self-energy must be negative everywhere in the upper complex plane. This combined with the Kramers-Kronig relationships results results in the identity

\begin{align}\label{prop_S2}
\Sigma(z) &=  \frac{1}{2\pi} \lim_{\delta\rightarrow0^+} \int_{-\infty}^{\infty} d\omega \frac{\rm{Im}\big[\Sigma(\omega+\imath\,\delta)\big]}{\omega-z}, \nonumber \\
&\lim_{\delta\rightarrow0^+}\:\rm{Im}\big[\Sigma(\omega+\imath\,\delta)\big] < 0
\end{align}

\section{Continuous pole expansion for the self-energy}\label{CPE}

Motivated by Eq.~(\ref{prop_S2}), the self-energy on the Matsubara frequencies can be parametrized with a positive, real function $f$,

\begin{align}\label{S_pade}
\tilde{\Sigma}(\varpi_m) =&  \: \Sigma_0 + \int_\infty^\infty d\omega \: \frac{f(\omega)}{\imath \, \varpi_m-\omega},\\
 \mbox{with}\quad f \geq& \: 0, \Sigma_0 \in \mathbb{R}, \quad \int_\infty^\infty dx \: f(x) = \Sigma_1. \nonumber
 \end{align}

The aim of the CPE-algorithm is to search for the positive function \textit{f}, that minimizes the norm $\Lambda$,

\begin{align}\label{norm_1}
\Lambda(\,\textit{f}\,)= \sum_{m=0}^{M} \Big| \tilde{\Sigma}(\varpi_m) - \Sigma(\varpi_m]\Big|^2 .
\end{align}

In order to find \textit{f}, we will decompose it in the basis of a regular spaced, piecewise linear function. If $\{\omega_n = n\: \Delta/N\}$ for $  n \in \{ -N, ..., N \}$ forms our regular spaced grid on the real axis, we can define the decomposition explicitly with the help of the step-function $\theta$,

\begin{align}\label{f_decomposition}
f(\omega)            =& \sum_{n=-N}^{N} \alpha_n \: \phi_n(\omega) \quad \mbox{with  } \alpha_n \geq 0 ,  \\
\phi_n(\omega) =& \: \theta(\omega-\omega_{n-1})\:\theta(\omega_n-\omega) \frac{\omega-\omega_{n-1}}{\omega_{n}-\omega_{n-1}} \nonumber \\
			&+ \: \theta(\omega-\omega_{n})\:\theta(\omega_{n+1}-\omega)  \frac{\omega_{n+1}-\omega}{\omega_{n+1}-\omega_{n}}. \nonumber
\end{align}
 
 \noindent
Due to this explicit decomposition, we can perform the integral in Eq.~(\ref{S_pade}) analytically and rewrite $\tilde{\Sigma}(z)$ into a much simpler form

\begin{align}
\tilde{\Sigma}(z) =& \: \Sigma_0 +  \sum_{n=-N}^N \Phi_{n}(z)\: \alpha_n \quad \mbox{with  } \alpha_n \geq 0 ,  \\
\Phi_{n}(z) =& \frac{\omega _{n-1}-z}{\omega _{n-1}-\omega _n}   \log \left(\frac{\omega _{n-1}-z}{\omega _n-z}\right) \nonumber \\
&\quad-\frac{\omega_{n+1}-z}{\omega _n-\omega _{n+1}} \log \left(\frac{z-\omega _n}{z-\omega _{n+1}}\right)  \nonumber
 \end{align}

Next we define a transfer matrix $A_{m,n}=\Phi_n(i\:\varpi_m)$ and rewrite the norm $\Lambda$ into a least square problem with boundary conditions,
 
\begin{align}
\Lambda &=  \sum_{m=0}^{M} \Big| \Sigma_0+\sum_{n=-N}^{N}A_{m,n}\,\alpha_n - \Sigma(\varpi_m)\Big|^2 , \quad\alpha_n \geq 0.  \nonumber
\end{align}

\noindent
By expanding the norm and gathering the terms of the same order in $\alpha_n$, one obtains an explicit quadratic form for the norm $\Lambda$ as a function of the constant transfer matrix $A_{m,n}$ and the coefficients $\Sigma_0$ and $\{\alpha_n\}$,
 
 \begin{align}\label{norm_2}
\Lambda &= M\:\Sigma_0^2 -2\:\Sigma_0\:\sum_{m=0}^{M-1} \mbox{Re}[\Sigma(i\:\varpi_m)] \nonumber \\
	&\quad +  \vec{\alpha}\:Q\:\vec{\alpha}^T +  \vec{q}\:\vec{\alpha}^T + C,\\
Q &= \Big(\mbox{Im}[A]^T \mbox{Im}[A] + \mbox{Re}[A]^T \mbox{Re}[A]\Big) \nonumber \\
\vec{q} &= 2\:\Big(\big(\Sigma_0-\mbox{Re}[\vec{\Sigma}]\big)\:\mbox{Re}[A] - \mbox{Im}[\vec{\Sigma}]\:\mbox{Im}[A] \Big)\, \nonumber \\
\vec{\alpha} &= \{\alpha_{-N}, ... \:, \alpha_N\} \mbox{  and  } \nonumber\\ \vec{\Sigma} &= \{\Sigma(i\varpi_0), ... \:, \Sigma(i\varpi_M)\}. \nonumber
\end{align}

The CPE algorithm thus rephrases the problem of analytic continuation of a noisy function to a quadratic programmable optimization problem, with linear constraints. These type of problems are well known, and many algorithms exists to find the minimum. From Eqs.~(\ref{norm_2}), it follows that Q is positive semi-definite and according to Eq.~(\ref{norm_1}), $\Lambda$ has a trivial lower bound. In a a quadratic programmable optimization problem, these conditions are sufficient to guarantee a unique solution for which our norm $\Lambda$ is minimized. Consequently, given $\Delta$, $N$, and $M$, there is a \textit{unique} set of $\{\alpha_n\}$ and $\Sigma_0$ that minimize the norm $\Lambda$. The Frank-Wolf algorithm\cite{FrankWolfe1956} (FWA) is the simplest for solving a quadratic programmable optimization problem. It is applied in the context of the CPE-algorithm in the following way:  First, we subtract the zeroth moment $\Sigma_0$ of the measured self-energy $\Sigma$. This ensures integrability of $\Sigma$ and $\tilde{\Sigma}$ along the real axis. Next, we choose an initial set of $\{\alpha\}$, and compute the gradient of $\Lambda$ towards $\{\alpha_n\}$ and $\Sigma_0$,

\begin{align}\label{grad_S_pade}
\vec{\nabla}_{\{\alpha\}}\Lambda &=  2 \: \mbox{Im}[A]^T   \Big(  \mbox{Im}[A] \vec{\alpha} - \mbox{Im}[\vec{\Sigma}]\Big) \nonumber \\
&\quad+ 2 \: \mbox{Re}[A]^T   \Big(  \mbox{Re}[A] \vec{\alpha} + \Sigma_0 -\mbox{Re}[\vec{\Sigma}]\Big). \nonumber \\
\frac{\partial \Lambda}{\partial \Sigma_0} &= 2\:\sum_{m=0}^{M-1} \Big(\mbox{Re}[A] \vec{\alpha} + \Sigma_0 - \mbox{Re}[\Sigma(i\:\varpi_m)]\Big)
\end{align}

\noindent
Notice that the initial guess of $\{\alpha_n\}$ is unimportant, since there is only 1 minimum in our convex search-space. We now search a $\lambda$, that minimizes the norm $\Lambda$ along the direction of $-\vec{\nabla}_{\{\alpha\}}\Lambda$. Special care has to be taken to enforce positivity of all coefficients $\{\alpha\}$. This is accomplished by point wise application of the $\rho$ ramp-function\footnote{\textit{http://mathworld.wolfram.com/RampFunction.html}}.

\begin{align}
\tilde{\Sigma}(\lambda) &= A \: \rho( \vec{\alpha} \: - \: \lambda \:\vec{\nabla}_{\{\alpha\}}\Lambda)\\
\Lambda(\lambda) &= \sum_{m=0}^{M} \Big( \mbox{Im}\Big[\tilde{\Sigma}(\lambda) - \vec{\Sigma}\Big]\Big)^2 + \Big( \mbox{Re}\Big[\tilde{\Sigma}(\lambda) - \vec{\Sigma}\Big]\Big)^2. \nonumber
\end{align}

The parameter $\lambda_{min}$ that minimizes our norm, can now be used to generate a new set of coefficients $\{\alpha\}$,

\begin{align}
\vec{a}_{i+1} &= \rho(\, \vec{\alpha}_i \: - \: \lambda_{min} \:\vec{\nabla}_{\{\alpha_i\}}\Lambda\,).
\end{align}

\noindent
We continue this iterative process until $\Lambda$ is \textit{numerically} converges to a minimum value. 

This minimization approach of the CPE-algortihm has several major benefits compared to other algorithms. First, the CPE-algorithm depends only on 3 \textit{external} parameters $\Delta$, $N$ and $M$, as we do not consider the to-be-fitted self-energy points on the imaginary axis as separate degrees of freedom. Furthermore, for these 3 parameters, there is a \textit{unique} solution, since the norm can be rewritten as a quadratic function with a positive semi-definite kernel-matrix $A$. Second, CPE is numerically stable against noise on the measured self-energy $\Sigma$, since we use a fitting procedure. This robustness is important in the DMFT context, where the self-energy is computed via a stochastic process. Third, CPE is a self-consistent method, which returns a measure for the quality of the fit via $\Lambda$. This measure can be used to adjust the \textit{external} parameters in the case of a bad fit.

\section{Validation of CPE with Exact Diagonalization.}\label{EDresults}

\begin{figure*}[!]
\begin{center}
\includegraphics[width=\textwidth]{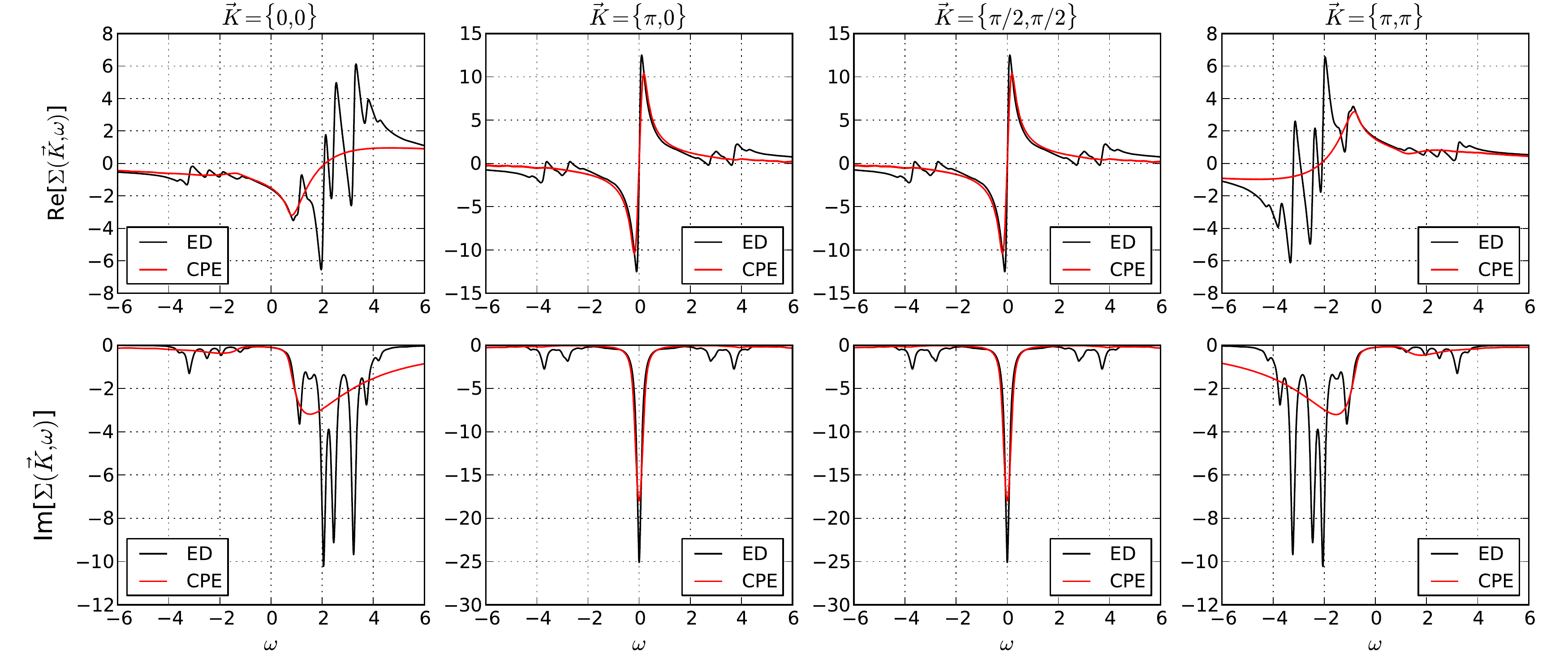}
\end{center}
\caption{\label{fig:ED} Comparison the real and imaginary part of the self-energy $\Sigma(\vec{K}, \omega)$ obtained with ED as well as with CPE. We observe that the CPE can capture the self-energy remarkably well around the Fermi-energy ($\omega=0$), as well as the broad features far away from it.}
\end{figure*}

Our first validation step is to compare the CPE algorithm to ED results for exactly solvable models. Without too much difficulties we can presently solve the single band Hubbard model on an isolated 8-site cluster. Due to the finite size of the system, the Hamiltonian has a limited and manageable number of terms and is represented as  

\begin{align}
\mathcal{H} = &-t  \sum_{\sigma=\uparrow, \downarrow} \sum_{\langle i,j \rangle = 1}^8 c_{i, \sigma}^{\dagger} \, c_{j, \sigma} \nonumber \\
		&+ \frac{U}{2} \sum_{\sigma=\uparrow, \downarrow} \sum_{i = 1}^8  (n_{i, \sigma}-1/2)(n_{j, -\sigma}-1/2) \nonumber
\end{align}

\noindent
The Hamiltonian acts on a Fock-space, composed of $2^{16}=65536$ states.  After applying total number and magnetization symmetries, the matrix can be block-diagonalized with a maximum block-size of $4900$. Using standard eigenvalue decomposition routines of LAPACK\cite{LAPACK}, we can obtain all the eigen-energies $\{\epsilon_i\}$ and eigenstates $\{\vert \Psi_i \rangle\}$ of the isolated 8-site cluster. Following standard many-body theory, we can now compute the Greens-function anywhere in the (upper) complex plane

\begin{align}
G(\nu, \mu, z) = \sum_{i,j} \frac{e^{-\beta\, \epsilon_i}}{\mathcal{Z}} \frac{\langle \Psi_i \vert c_{\nu} \vert \Psi_j \rangle \langle \Psi_j \vert c^{\dagger}_{\mu} \vert \Psi_i \rangle }{z-\epsilon_i+\epsilon_j}. 
\end{align}

\noindent
Here, the symbols $\nu$ and $\mu$ are short-hand notations for the band, spin and cluster K-point $\nu = \{b_{\nu}, s_{\nu}, \vec{K}_{\nu}\}$. By solving the cluster twice, once with and once without the interaction, we can obtain, respectively, the interacting and non-interacting Greens-function. From these two functions, we can obtain the self-energy in momentum space through the Dyson's equation anywhere in the complex plane,

\begin{align}
\Sigma(\vec{K}, z) = G_0^{-1}(\vec{K}, z)-G^{-1}(\vec{K},z).
\end{align}

\noindent
The idea is now to evaluate the self-energy on the real axis and on the Matsubara frequencies that are located on the imaginary axis. In this way, we can use the  self-energy on the Matsubara frequencies as an input for the CPE, and compare the analytically continued self-energy with the exact result.

In Fig.~\ref{fig:ED}, we show the self-energy for various $\vec{K}$-points on the real axis , with an off-set of $\delta=0.1\imath$. By comparing the ED results with the analytically continued self-energy, we can study the strengths and weaknesses of the CPE. Looking at the imaginary part of the self-energy, we observe that the high-frequency behavior of the self-energy decays smoothly and does not diverge. Furthermore, we see that the CPE captures remarkably well the self-energy around the Fermi-energy ($\omega=0$), as well as the broad features far away from it. However, sharp features are not reproduced. This is clear from the imaginary part of the self-energy at $\vec{K}=\{\pi,0\}$. The divergence at the Fermi-energy is underestimated and the features at the interval $\omega=[-4, -2]$ are absent in the CPE self-energy.

\begin{figure}[!]
\begin{center}
\includegraphics[width=0.5\textwidth]{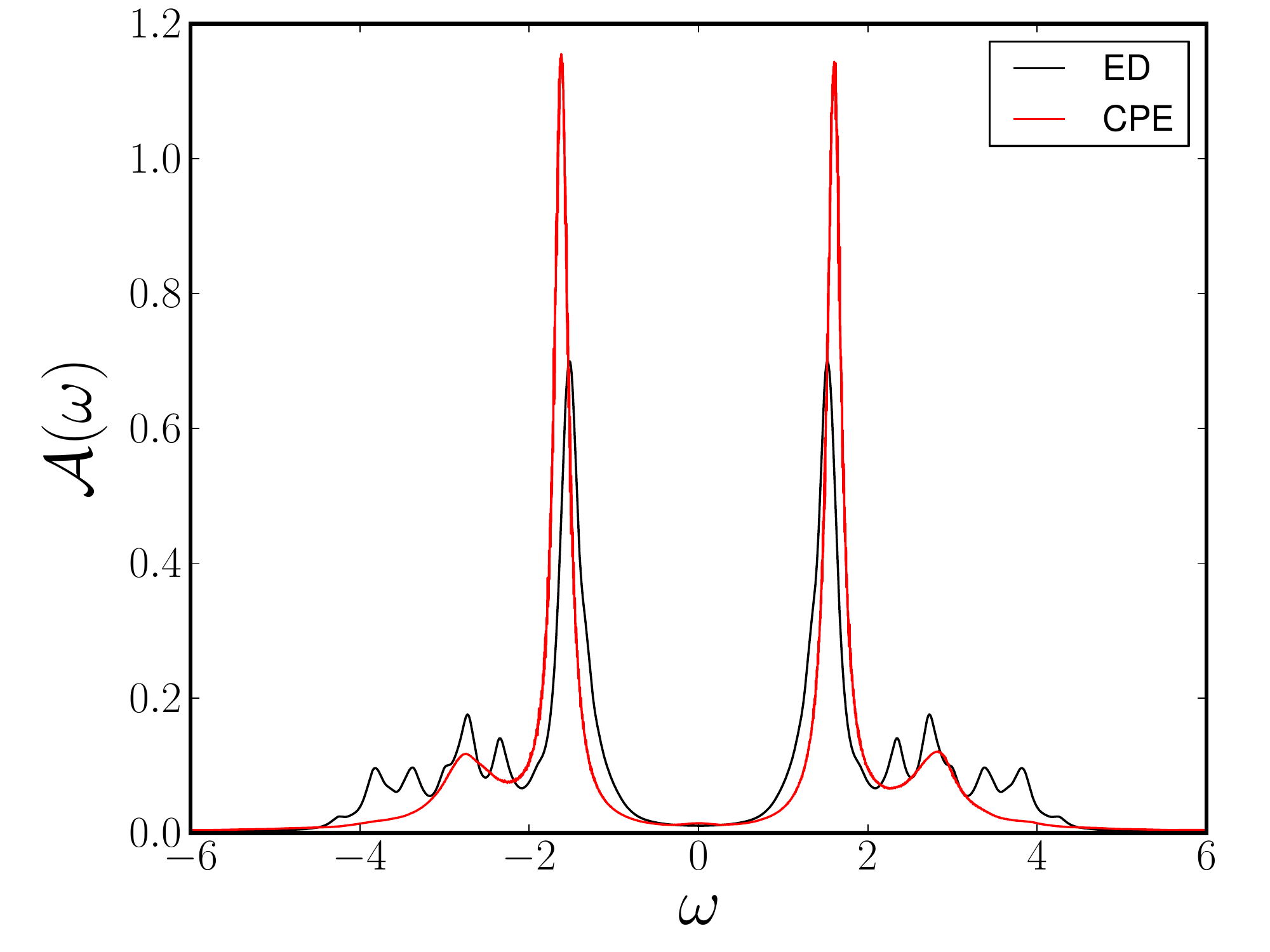}
\end{center}
\caption{\label{fig:ED_2} Comparison of the spectrum  obtained with ED as well as with CPE. One can clearly observe that the gap around the Fermi-surface is well represented by the CPE, as well as the broad Hubbard-bands at $\omega \approx \pm3$.}
\end{figure}

Consequently, one should not expect to be able to determine sharp features in the spectrum with the CPE far from the Fermi-surface. Only broad features are captured accurately far from the Fermi-surface. This is illustrated by Fig.~\ref{fig:ED_2}, where we compare the spectra obtained with ED and CPE, respectively. One can see that the gap around the Fermi-surface is well represented by the CPE, as well as the broad Hubbard-bands at $\omega \approx \pm3$.

The spectrum of lattice models is generally smoother than that of finite size clusters, since they have an infinite number of eigenvalues instead of a finite set. The finite set of eigenvalues introduces poles with a finite weight on the real axis and thus gives rise to the sharp features observed in Fig.~\ref{fig:ED} and Fig~\ref{fig:ED_2}. Consequently, we expect that the CPE should perform better for lattice models than for the finite size model we just considered.

\section{Application to lattice problems}\label{results}

We now apply the CPE algorithm to two well known lattice problems, in order to demonstrate the ability of the algorithm to reproduce published numerical or experimental results. We will consider two fundamentally different problems. First we use the CPE to investigate the momentum dependence of the spectrum in the single-band Hubbard model in two dimensions at half filling. Particularly the momentum dependence of the gap formation has received a lot of interest recently\cite{Werner2009PRB, Gull2010PRB}, and there are several results in the literature to compare to. We can further validate the CPE by looking at the $\vec{K}$-dependent self-energy obtained from quantum Monte Carlo simulations within the DCA.  Second, we will use the CPE to compute the spectrum of NiO and compare the latter to experimental data. The prediction of a 4.3 eV gap around the Fermi-energy is one of the great successes of LDA+DMFT, and consequently has to be reproduced by CPE based spectral functions. We will show that this is the case, and that in combination with LDA+DMFT, the CPE can be seen as a practical and unambiguous tool to compute spectral functions of real materials that compare rather well with experiments.

\subsection{Momentum-dependent gap formation in half-filled 2D Hubbard model.}

\begin{figure}[!]
\begin{center}
\includegraphics[width=0.5\textwidth]{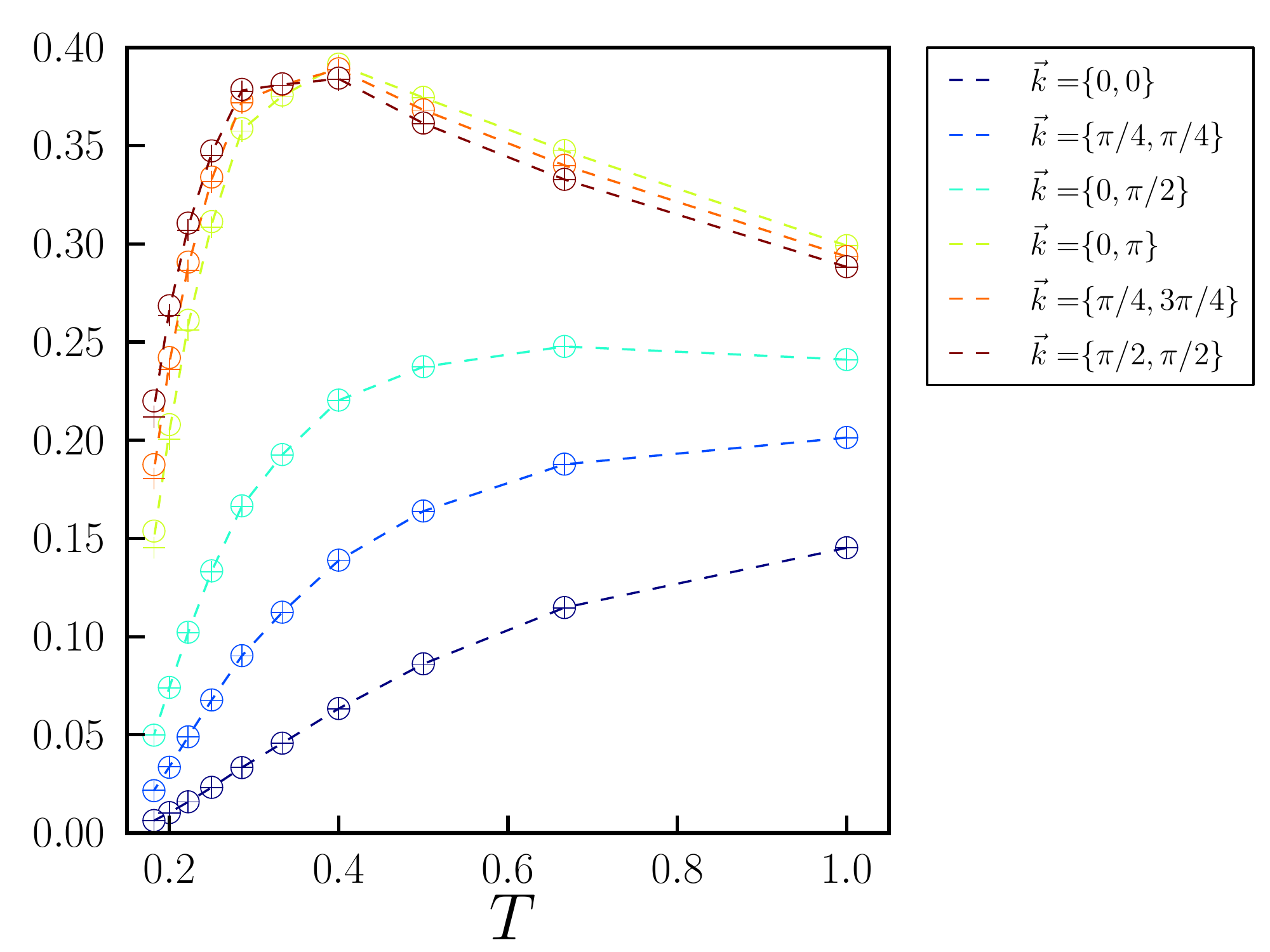}
\end{center}
\caption{\label{fig:verification}. Verification of the CPE algorithm through Eq.~(\ref{cheating}) at zero percent doping. The left hand side $\beta\,G(\vec{k},\beta/2)$ is depicted by crosses, while the right hand side is given by the solid dots. The dotted line connects the average of the left-hand side and right hand side. Given that this relationship is not enforced during the minimization-process, the agreement is remarkably good.}
\end{figure}

The combination of the Dynamical Cluster Approximation together with the CPE algorithm allows us  to investigate the momentum dependency of the spectrum in different regimes of the phase-diagram. Recently, much attention has been given to the momentum dependent gap formation at half-filling in the single band Hubbard model. In particular, it has been shown\cite{Werner2009PRB,Gull2010PRB} that for a specific interaction-strength of $U/t=6$, the anti-nodal regions ($[\pm\pi,0]$ and $[0,\pm\pi]$) lose their spectral weight contribution at the Fermi-energy faster than the nodal regions ($[\pm\pi/2,\pm\pi/2]$). This momentum anisotropy in the self-energy is a very interesting phenomenon since exotic ground-states such as the anti-ferromagnetic and d-wave superconducting state require a momentum dependent self-energy. 

\begin{figure}[!]
\begin{center}
\includegraphics[width=0.5\textwidth]{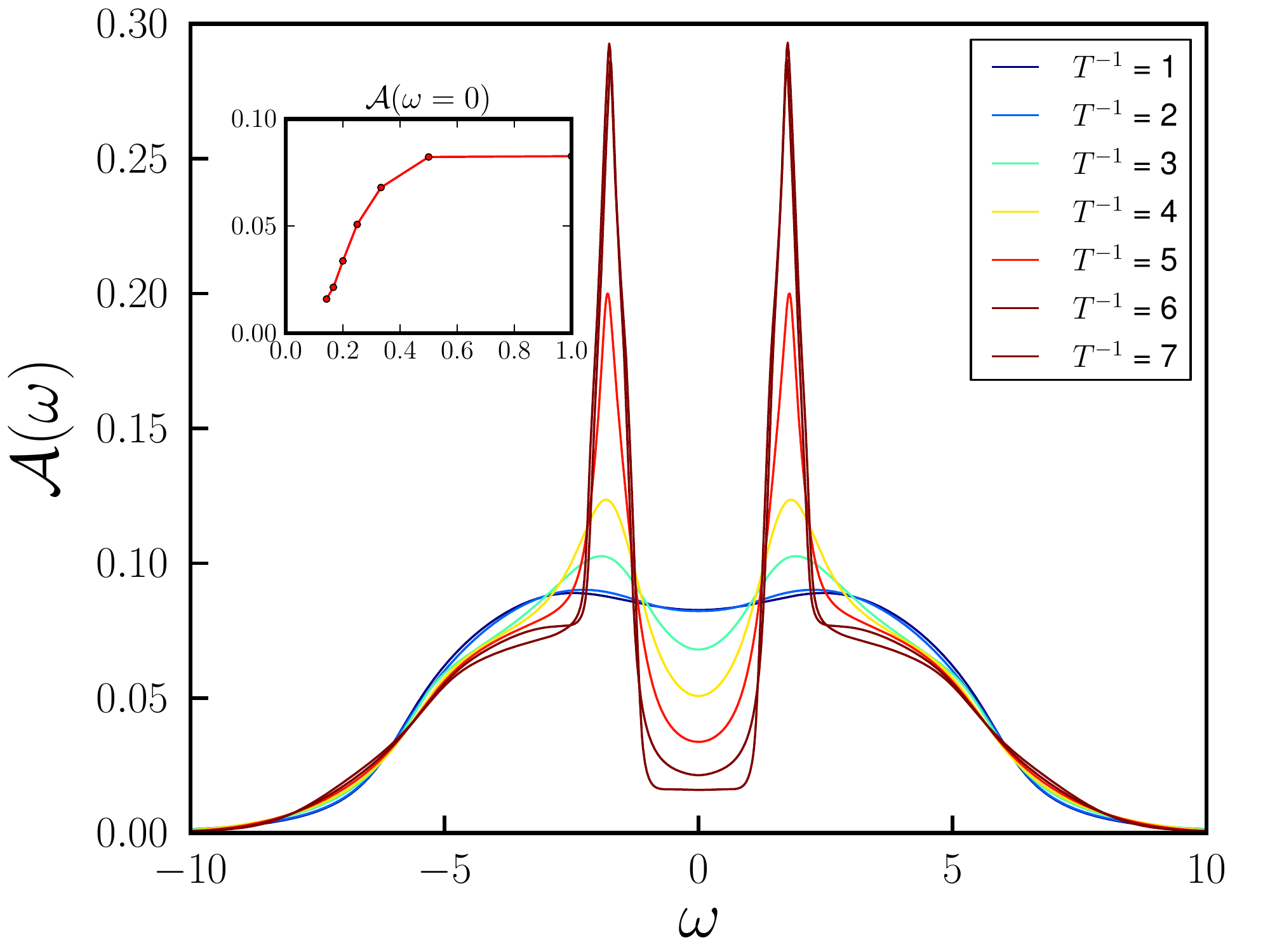}
\end{center}
\caption{\label{fig:half_filling} Temperature dependence of the spectrum $\mathcal{A}$ at half-filling. Notice the appearance of the van-Hove singularities originating from the band splitting at the Fermi-energy $\omega=0$.}
\end{figure}

\begin{figure*}
\begin{center}
\includegraphics[width=\textwidth]{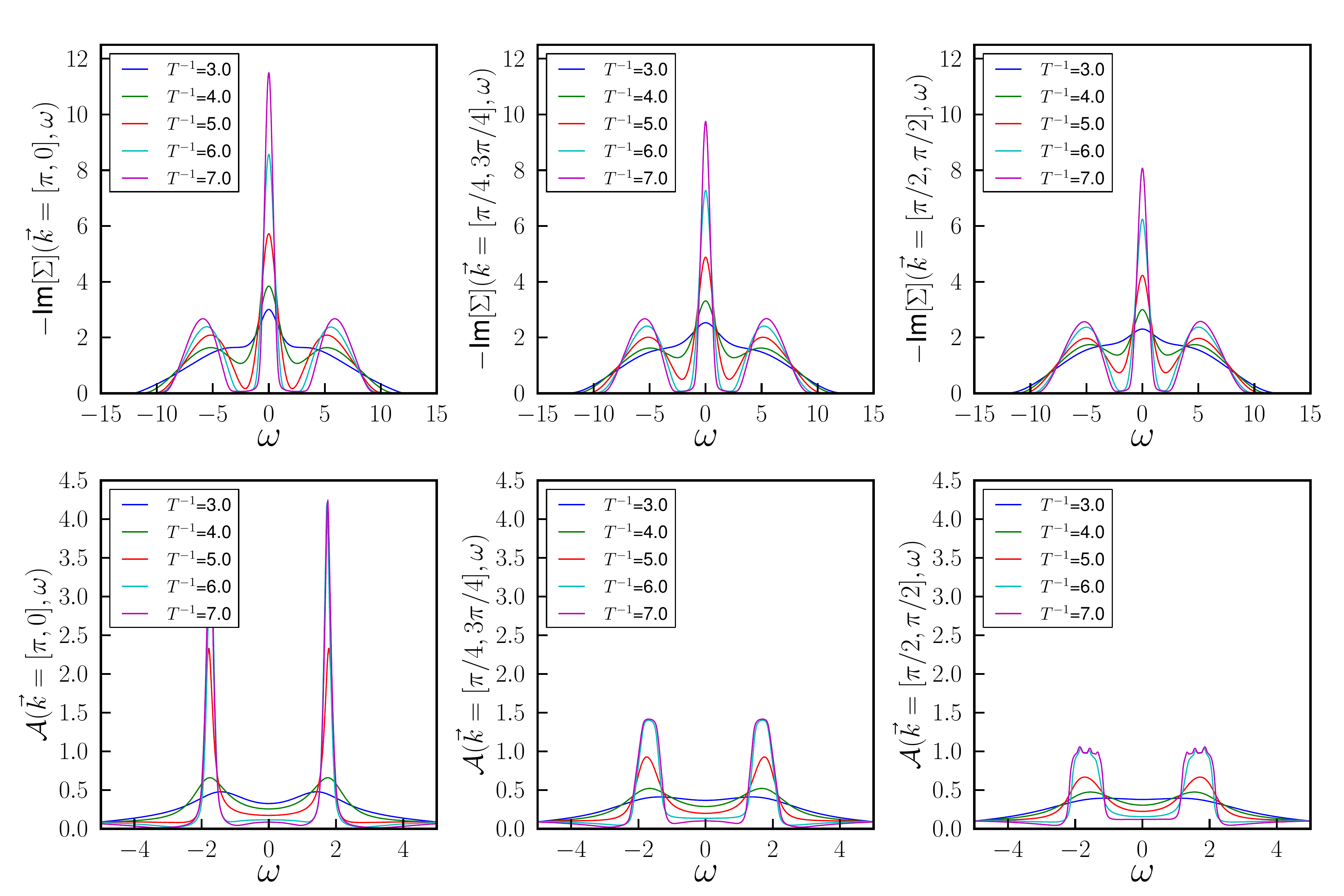}
\end{center}
\caption{\label{fig:mom_anis_half_filling} Temperature and momentum space dependence of the self-energy and spectrum along the Fermi-surface in the single band Hubbard model at half filling.}
\end{figure*}

Here we investigate this momentum anisotropy on a 32-site cluster at half filling for an interaction strength of $U/t=6$. A 32-site cluster is large enough to allow a careful study of the momentum-anisotropy along the Fermi-surface, and will give us an intimate view on the gap-formation at zero doping. In the literature, the momentum anisotropy of the gap formation has been investigated by using the identity, 

\begin{align}\label{cheating}
\beta \, G(\vec{k}, \tau = \beta /2 ) &= -\beta/2 \int d\omega \frac{\mathcal{A}(\vec{k},\omega)}{\cosh(\beta \omega /2)}\\
										  &\stackrel{\beta \rightarrow \infty}{\approx} \mathcal{A}(\vec{k},\omega=0). \nonumber
\end{align}

At low temperatures, the quantity $\beta \, G(\vec{k}, \tau=\beta /2 )$ provides a good estimate for the spectrum at the Fermi-energy, since the function $1/\cosh(\beta \omega /2)$ becomes a delta-function for $\beta$ going to infinity. Since the imaginary time function can be computed directly with a quantum Monte Carlo solver, the spectrum at the Fermi-energy can be probed straightforwardly, without the need to do an analytical continuation of the self-energy or Greens-function. With the CPE, we can obtain the spectrum $\mathcal{A}(\vec{K}, \omega)$ on the entire real axis. We can thus verify the accuracy of the CPE by performing the integral on the right-hand side of Eq.~(\ref{cheating}) and comparing to the quantity $\beta \, G(\vec{k}, \tau=\beta /2 )$.

The crosses in Fig.~\ref{fig:verification} show for various $\vec{k}$-points the left hand side of Eq.~(\ref{cheating}). The open circles are the results for the corresponding right hand side of the equation, where the integral was computed with the CPE (note that equality is not enforced during the minimization process). The agreement is very good and implies that all features we compute with the CPE-algorithm at or near the Fermi energy will be in agreement with results in the literature that have been computed with the left hand side of Eq.~(\ref{cheating}).

The Mott-transition at zero percent doping has been intensively investigated since the first DMFT results became available\cite{Georges1996RMP, Maier2005RMP}. The behavior of the spectrum as a function of the temperature is therefore well known in this region and offers a benchmark the CPE algorithm. In Fig.~\ref{fig:half_filling}, we show the temperature dependence of the spectrum $\mathcal{A}(\omega)$. As expected, the spectrum drops quickly in a broad region around the Fermi-energy ($\omega=0$) in order to form a gap. At the same time, two broad Hubbard bands emerge at $\omega=\pm4$ and two sharp peaks emerge at the edge of the gap. These sharp features originate from the van-Hove singularities, which typically go together with the formation of a gap. The advantage of analytically continuing the self-energy as opposed to the Greens-function now becomes clear. Since the CPE reliably reproduces broad features it is perfectly well suited for analytic continuation of irreducible quantities like the self-energy, which are assumed to be rather smooth. The sharper features such as the van-Hove singularities will then be generated by the Brillouin-zone integration of the Greens function on the real axis:

\begin{align}\label{spectrum_integration}
\mathcal{A}(\omega) = -\frac{1}{\pi} \int \, d\vec{k} \: \rm{Im}\Big[\frac{1}{\omega + \imath \, \delta + \mu - \epsilon(\vec{k}) - \Sigma(\vec{K},\omega)}\Big]
\end{align}

\noindent
Here, we used a common off-set of $\delta=0.1$. The Brillouin zone integration is performed using the tetrahedron integration method\cite{Lambin1984PRB,Lee2002PRB} (TIM). The TIM was developed especially to handle integrals over inverse functions. The inversion introduces poles in the integrand, and TIM can treat these poles in a numerically controlled way. 

\noindent
Next we focus on the momentum anisotropy of the self-energy, and its impact on the spectrum. In Fig.~\ref{fig:mom_anis_half_filling}, we show the imaginary part of the self-energy and the spectrum for three different $\vec{k}-$points along the Fermi-surface. By looking at the rate of divergence on the imaginary axis of the self-energy and investigating the local densities, Werner et al.~\cite{Werner2009PRB} argued that the spectrum at the Fermi-energy on the anti-nodal points should disappear at a much faster rate then at the nodal points. The CPE confirms these findings. The imaginary part of the self-energy at the anti-nodal points is much larger than at the nodal points. Since the spectrum is inversely proportional to the self-energy, the spectrum vanishes faster at the anti-nodal points.  Furthermore, we can immediately observe that this anisotropy in the self-energy increases as the temperature $T$ is lowered. Using the partial occupancies $n_{\vec{K}}$ in the different patches of the Brillouin zone, Gull et al.~\cite{Gull2010PRB} have claimed that the gap opened at the nodal region should be much bigger than at the anti-nodal region. Defining the width of the gap as the minimum distance between the two van Hove singularities in Fig.~\ref{fig:mom_anis_half_filling}, it can be seen that the claim of an anisotropic gap in the spectrum is also confirmed. 

\noindent
Finally we discussed features which to our knowledge have not yet been reported in the literature. In particular, we would like to draw the attention to the formation of the valleys in the imaginary part of the self-energy at $\omega \approx \pm 2$. Going from $T^{-1}$=3 to $T^{-1}$=7, we see that these valleys grow faster and are more profound at the anti-nodal points than at the nodal points. Since the imaginary part of the self-energy can be thought of as the inverse lifetime of the quasiparticle, we can conclude that the quasiparticle will have a short lifetime on the Fermi-energy, and a much longer one in the valleys. This picture translates directly into to the spectrum, where two peaks rapidly grow at $\omega \approx \pm 2$. The difference in the shape of these peaks can be explained by the topology of the free dispersion $\epsilon(\vec{k})$. The free dispersion is extremely flat at the anti-nodal points, since both the first derivative and the laplacian vanishes at this point). The spectral weight at the nodal point will thus be extremely peaked at the Fermi-energy if no interaction is present. However, if there is a non-zero interaction, the lifetime of the particles at the Fermi-energy will be very short, and all of them will be scattered in equal amount to higher of lower energy-levels. At the nodal points, the free dispersion spectrum is essentially linear with Fermi-velocity vector $\epsilon(\vec{k}) \sim \vec{k} \, \vec{v}_F $). Hence, the free spectrum will be smeared around the Fermi-energy and have some spectral weight in the valleys of the the imaginary part of the self-energy. Hence, these electrons will not be scattered away, when an interaction is introduced into the system. With this simple picture in mind, one can now easily understand shape-difference of the peaks in the spectrum shown in Fig.~\ref{fig:mom_anis_half_filling}, as well as the findings of Gull et al. with the partial occupancies as a function of the chemical potential.

\subsection{Electronic structure of NiO}\label{NiOsec}

\noindent
We will now apply the CPE to a multi-band model of NiO with materials specific parameters derived from first principles electronic structure calculations. NiO is a prototypical material with strong electronic correlations that has been extensively studied both experimentally~\cite{McKay1984PRL,Sawatzky1984PRL,Shen1991PRB} and theoretically~\cite{Terakura,ren2006lda+,Kunes2007PRB,Kunes2007PRL}. The experimental results from the literature for this compound are the next best thing compared to the exact solution of the impurity model of a multi-band system. It is thus an ideal testing ground for electronic structure calculations. Here we generate the Green's function and self-energies from $\rm{LDA}+\rm{DMFT}$ ~\cite{Kotliar2006RMP,Karolak2010JESRP} based Monte Carlo simulations, and subsequently apply the CPE for analytic continuation of the self-energy to obtain spectral function, which we compare directly to experiment. 

The large insulating band gap of 4.3 eV can not be predicted by conventional band theory. Density Functional Theory within the Local Spin-Density Approximation (LSDA)\cite{Terakura} predicts NiO to be a band insulator, where the band gap is consequence of the anti-ferromagnetic order. Angle-resolved photoemission experiments (ARPES), however, have shown~\cite{Tjernberg1996PRB} that the electronic band gap also exists in the paramagnetic phase, far above the Neel temperature at $525K$. Furthermore, the predicted band-gap with LSDA is considerably smaller than the experimental values. These deficiencies are corrected for by introducing correlation effects via DMFT. Here we will compare our calculated $\rm{LDA}+\rm{DMFT}+\rm{CPE}$ electronic spectra with experimental spectroscopy measurements in order to further validate the CPE method.

\begin{figure}[!]
\begin{center}
\includegraphics[width=0.5\textwidth]{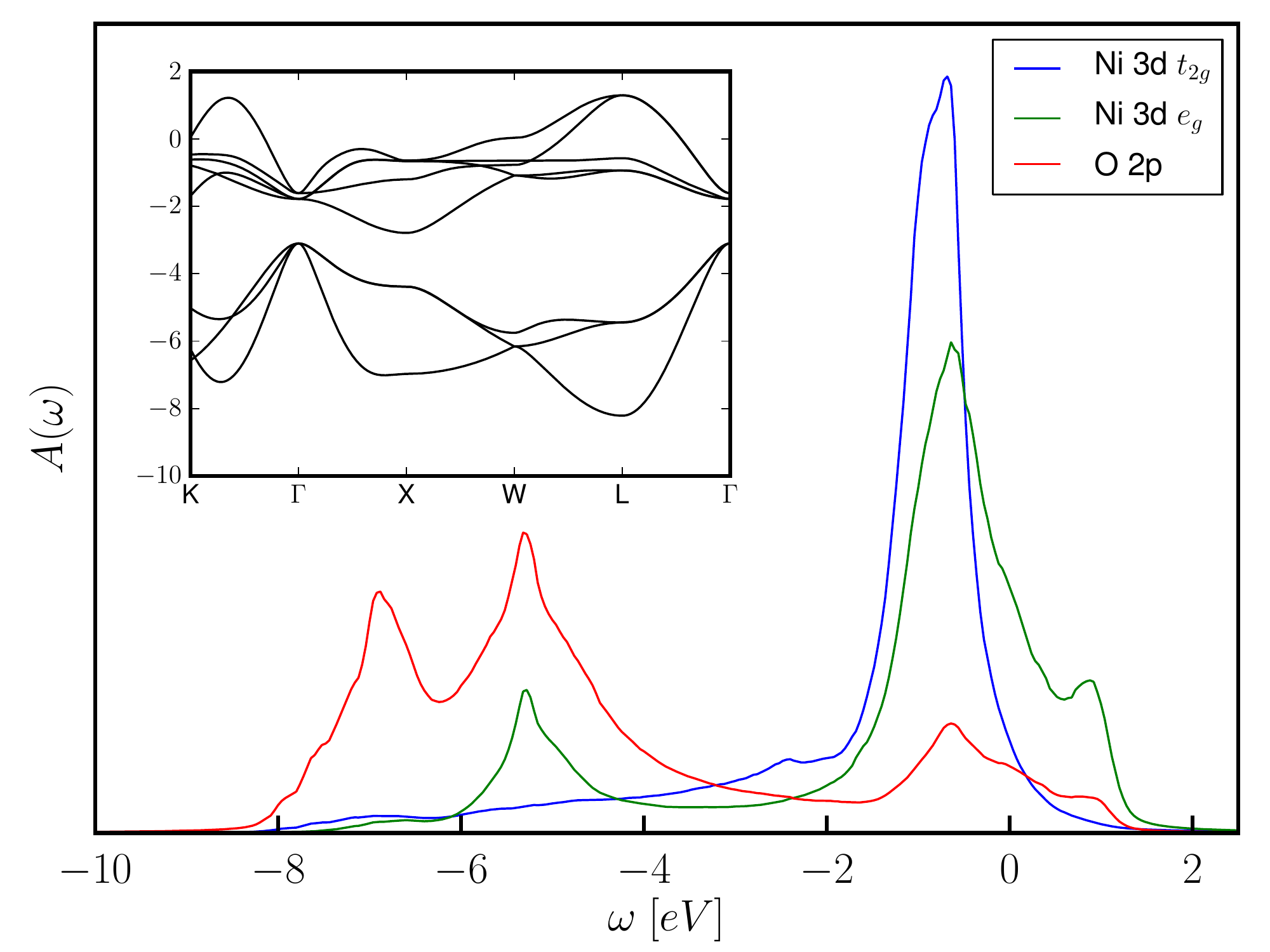}
\caption{The non-correlated partial density of states of NiO as obtained by LDA. Inset: band-structure of NiO obtained with LDA.}
\label{fig:NiO_LDA} 
\end{center}
\end{figure}

The impurity Hamiltonian $\mathcal{H}$ we study here is given in Eq.~\ref{NiO_Hamiltonian}. It is of the usual LDA+U form. In the present study, the LDA band-structure of NiO was obtained by all-electron calculations using the linearized augmented plane wave (LAPW) method and is displayed in the inset of Fig.~\ref{fig:NiO_LDA}. The resemblance of the band-structure reported by Karolak et al. ~\cite{Karolak2010JESRP} is perfect and we can clearly observe the five Nickel-bands around the Fermi-energy $\omega=0$ as well as the lower three Oxygen-bands around $\omega \approx -6$. The interaction tensor $ U_{\nu,\sigma, \mu, \sigma'}$ was obtained through a constrained RPA calculation (c-RPA)~\cite{Kozhevnikov2010SC}. As is usual in the NiO compound, we have only kept the interaction-terms between the Nickel-orbitals and ignore the density-density interaction between the Oxygen-Oxygen and Oxygen-Nickel orbitals. In the c-RPA method, the Wannier-orbitals, which are used to construct the tight-binding Hamiltonian $\mathcal{H}_{\rm{LDA}}$ of the impurity, are reused to construct the interaction tensor $ U_{\nu,\sigma, \mu, \sigma'}$. As a consequence, the interaction terms are more consistent with the band-structure than a simple application of the rotationally invariant Slater-Kanamori~\cite{Kanamori1963PTP,Slater1936PR} on-site interaction-matrix, which is traditionally constructed with the help of the parameters $U$ and $J$. However, a least squares fit of the the c-RPA matrix towards the parameters $U$ and $J$ reveals that the interaction-matrix can be approximated quite well with the parameters $U=9.14$ and $J=0.71$. These values do not differ tremendously from the original $U=8$ and $J=1$ parameters by Karolak. For the sake of completeness, we have listed the interaction tensor $U_{\nu,\sigma, \mu, \sigma'}$ in table~\ref{tab:NiO_interaction}. 

\noindent
The multi-band impurity problem in the self-consistent DMFT-loop was solved by an implementation of the CT-HYB algorithm~\cite{Werner2006PRL,Werner2006PRB}.  As is common in the literature\cite{Kunes2007PRL}, only the diagonal elements of the self-energy matrix are computed with the CT-HYB algorithm and the off-diagonal elements are ignored. Our calculations where performed at an inverse temperature of $\beta = 5 \, \rm{eV}^{-1}$. At this temperature, the material is in the paramagnetic state and the correlations are strong enough to introduce a band-gap. Since DMFT introduces correlations that have already been partly accounted for in the LDA functional, a double counting correction $\mathcal{H}_{\rm{dc}}$ needs to be applied our impurity-Hamiltonian $\mathcal{H}$ in Eq.~\ref{NiO_Hamiltonian}. For this we follow the standard procedure of Karolak et al.~\cite{Karolak2010JESRP}. It should be noted that the sum over the $m$ in the double-counting term $\mathcal{H}_{\rm{dc}}$ only runs over the Nickel orbitals.

\begin{align}\label{NiO_Hamiltonian}
\mathcal{H} = \mathcal{H}_{\rm{LDA}} - \underbrace{\mu_{dc} \sum_{m \sigma} n_{m \sigma}}_{\mathcal{H}_{\rm{dc}}} + \underbrace{\frac{1}{2}\,\sum_{\nu,\sigma, \mu, \sigma'} U_{\nu,\sigma, \mu, \sigma'} n_{\nu,\sigma}\, n_{\mu,\sigma'}}_{\mathcal{H}_{\rm{int}}}. 
\end{align}

\begin{table}[t]
\begin{center}
\begin{tabular}{l|ccccc}
$U_{\nu,\sigma, \mu, -\sigma}$ & $n_{\rm{t_{2g}},-\sigma}$& $n_{\rm{t_{2g}},-\sigma}$& $n_{\rm{e_{g}},-\sigma}$& $n_{\rm{t_{2g}},-\sigma}$& $n_{\rm{e_{g}},-\sigma}$ \\\hline \\
$n_{\rm{t_{2g}},\sigma}$& 9.14&7.60&7.37&7.60&8.28\\
$n_{\rm{t_{2g}},\sigma}$& 7.60&9.14&8.06&7.60&7.60\\
$n_{\rm{e_{g}},\sigma}$& 7.37&8.06&9.14&8.06&7.37\\
$n_{\rm{t_{2g}},\sigma}$& 7.60&7.60&8.06&9.14&7.60\\
$n_{\rm{e_{g}},\sigma}$& 8.28&7.60&7.37&7.60&9.14\\
\end{tabular}
\end{center}
\begin{center}
\begin{tabular}{l|ccccc}
$U_{\nu,\sigma, \mu, \sigma}$ & $n_{\rm{t_{2g}},\sigma}$& $n_{\rm{t_{2g}},\sigma}$& $n_{\rm{e_{g}},\sigma}$& $n_{\rm{t_{2g}},\sigma}$& $n_{\rm{e_{g}},\sigma}$ \\
 \hline \\
$n_{\rm{t_{2g}},\sigma}$& 0.00& 6.83& 6.49& 6.83& 7.85\\
$n_{\rm{t_{2g}},\sigma}$&6.83&	 0.00& 7.51& 6.83& 6.83\\
$n_{\rm{e_{g}},\sigma}$&6.49&	 7.51& 0.00& 7.51& 6.49\\
$n_{\rm{t_{2g}},\sigma}$&6.83&	 6.83& 7.51& 0.00& 6.83\\
$n_{\rm{e_{g}},\sigma}$&7.85&	 6.83& 6.49& 6.83& 0.00\\
\end{tabular}
\end{center}
\caption{\label{tab:NiO_interaction} The interaction tensor $U_{\nu,\sigma, \mu, \sigma'}$ for NiO obtained by c-RPA. The tensor can be approximated by the rotationally invariant Slater-Kanamori interaction-matrix, using  $U=9.14$ and $J=0.71$, which are close to the commonly accepted values of $U=8$ and $J=1$ in the literature.}
\end{table}

\noindent
No rigorous derivation is known for the double counting term $\mathcal{H}_{\rm{dc}}$. Furthermore, the Mott-insulator gap increases with decreasing value of the the double counting correction, and the parameter $\mu_{\rm{dc}}$ is thus tuned to fit experiment. Since we are only interested in validating the CPE algorithm, we apply the commonly used value $\mu_{\rm{dc}}=25\,\rm{eV}$ and test wether our method indeed reproduced a bad gap that is in agreement with the expected value of $4.3 \rm{eV}$.

\begin{figure}[t]
\begin{center}
\includegraphics[width=0.5\textwidth]{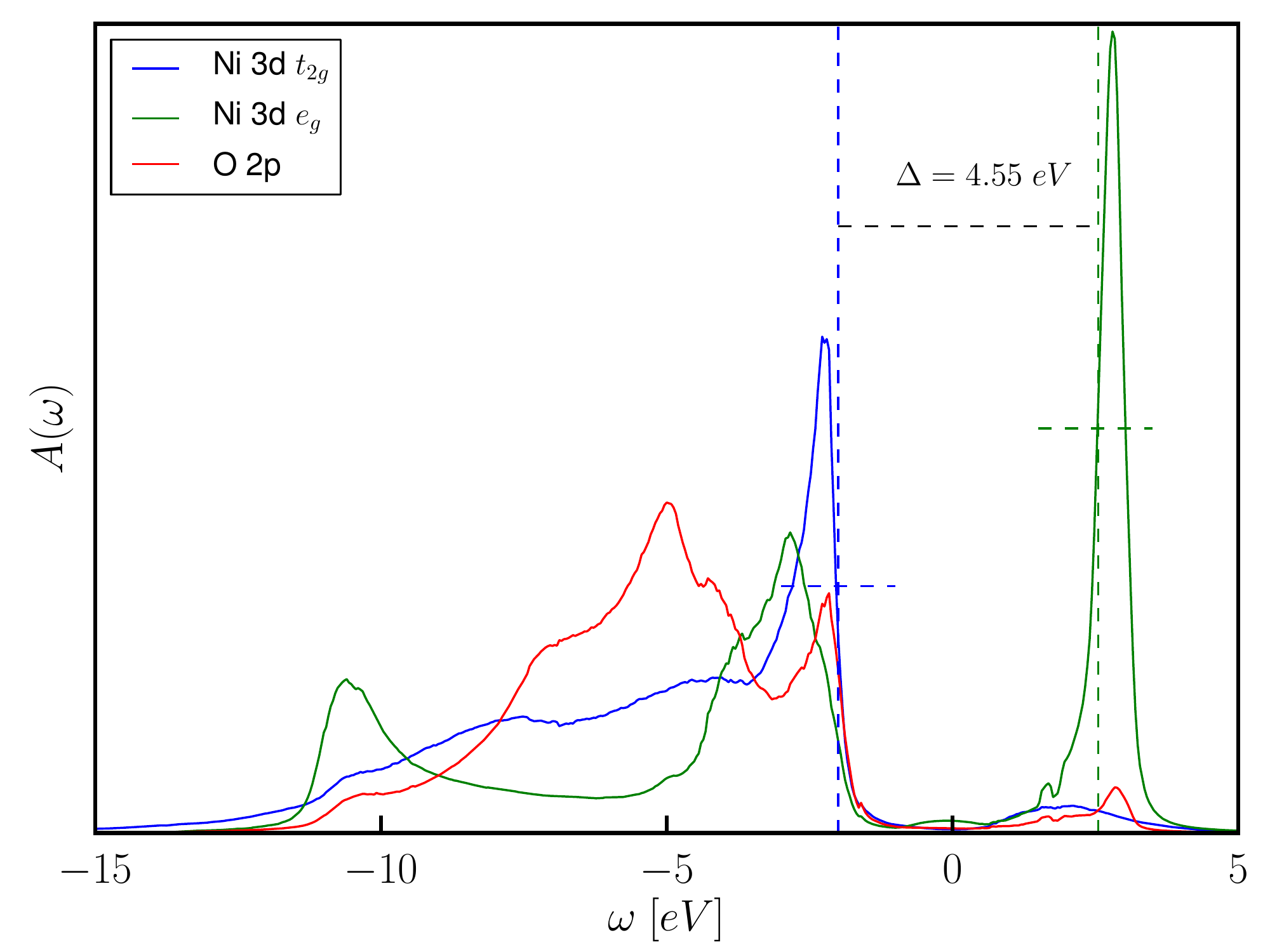}
\caption{\label{fig:NiO_DMFT} The correlated partial spectra of NiO calculated by $\rm{LDA}+\rm{DMFT}+\rm{CPE}$. Defining the band-gap as in experimental physics, i.e. the distance between the mid-points of the top of the peaks, we obtain a band-gap of 4.55 eV, in good agreement with the experimental value of 4.3 eV.}
\end{center}
\end{figure}

Once a density-density interaction is included between the Nickel-orbitals by means of a self-consistent DMFT calculation, a band gap appears in the spectral density. This can be clearly seen in Fig.~\ref{fig:NiO_DMFT}, where the partial spectra of each orbital are displayed. As usual, Fig.~\ref{fig:NiO_DMFT} was obtained by performing an analytical continuation of the Matsubara self-energy to the real axis with an off-set of $\delta=0.1$. A consecutive tetrahedron integration over the entire Brillouin zone then results in the lattice Greens-function, from which the partial spectrum $A_{\nu}(\omega)$ can be obtained. If we use the same definition of the band-gap from experimental physics, i.e. the distance between the mid-points of the top of the peaks, we obtain a band-gap of 4.55 eV, in good agreement with the experimental value of 4.3 eV obtained by Sawatsky and Allen~\cite{Sawatzky1984PRL}. These mid-points are represented by the horizontal dotted lines in Fig.~\ref{fig:NiO_DMFT}. Their intersection with the peak is marked by the vertical dotted lines, which difference defines the band-gap.

\begin{figure}[t]
\begin{center}
\includegraphics[width=0.5\textwidth]{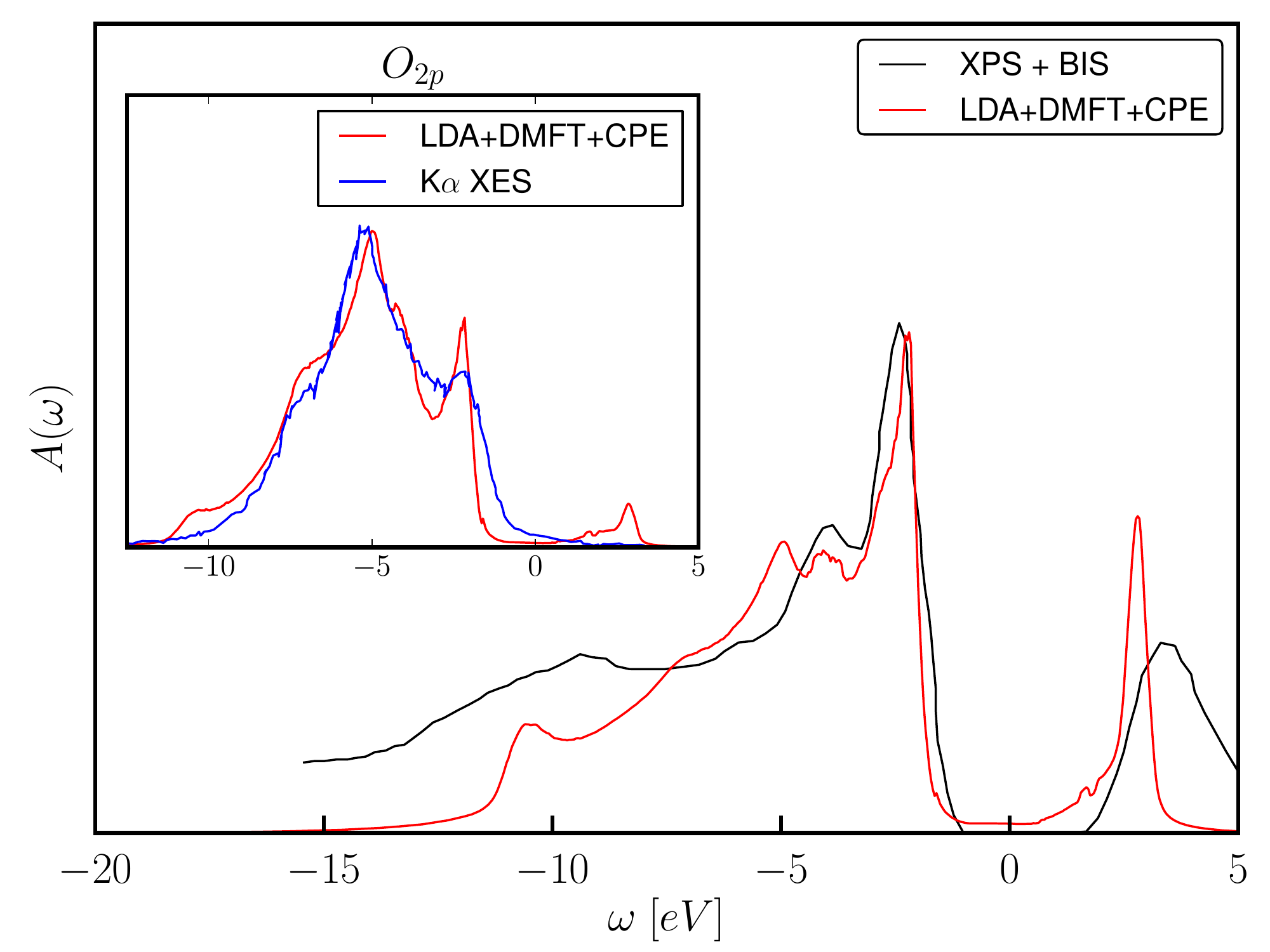}
\caption{\label{fig:NiO_XPS} The comparison of XPS and BIS spectra\cite{Sawatzky1984PRL} to the spectral function of NiO calculated by $\rm{LDA}+\rm{DMFT}+\rm{CPE}$. The XPS spectrum is measured at $120\,\rm{eV}$, showing primarily the Nickel 3d character. Inset: The comparison of K$\alpha$-emission spectra\cite{Kurmaev2008PRB} with the partial spectrum of the $O_{2p}$ orbital calculated by $\rm{LDA}+\rm{DMFT}+\rm{CPE}$.}
\end{center}
\end{figure}

From the comparison between the ED-results and CPE in section ~\ref{EDresults}, we know that the CPE is very good at reproducing an accurate picture for the spectrum close to the Fermi-energy. As such, it is not surprising to reproduce the correct band-gap of NiO around the Fermi-energy. To further bench-mark the CPE, we will compare the calculated spectrum with the experimental spectrum obtained by Sawatzky and Allen~\cite{Sawatzky1984PRL}. In this way, we want to explore how the CPE behaves over the entire real-axis and whether it can capture the essential physics far from the Fermi-energy, as we claimed in the Exact Diagonalization section. The experimental spectrum was obtained as a combination of X-ray photoemission (XPS) and Bremsstrahlung-Isocromat-Spectroscopy (BIS) measurements on cleaved single crystals of NiO. The XPS spectrum was recorded at $120\,\rm{eV}$ and mainly captures the Nickel 3d character. The measured spectra is shown in Fig.~\ref{fig:NiO_XPS}, together with the $\rm{LDA}+\rm{DMFT}+\rm{CPE}$ spectrum. The latter is obtained by summing up the partial spectra depicted in Fig.~\ref{fig:NiO_DMFT} and multiplied with the multiplicity of each orbital ($3\times t_{2g}, 2\times e_g, 3\times O_{2p}$). Since the spectroscopy is measured in arbitrary units of intensity, we can scale the measured spectrum such that the largest peaks have the same height. A simple comparison of both spectra shows a very good agreement between measured and computed spectrum. In the region $[-15, -5]$, we can observe that the CPE gradually rises and appears to reproduce some of the peaks, albeit with a slight left-shift of approximately $2\,\rm{eV}$. The peaks that define the gap around the Fermi-surface are also much sharper defined with the CPE. 

To further validate the CPE, we compare the partial spectrum of the $O_{2p}$ orbitals. The latter was measured very accurately by Kurmaev et al.\cite{Kurmaev2008PRB} with X-ray emission spectroscopy (XES). The Oxygen K-edge emission spectrum provides a representation of the $O_{2p}$ spectrum, and can thus be readily used to compare with the calculated $O_{2p}$ spectrum. In the inset of Fig.~\ref{fig:NiO_XPS}, we compare the measured with the computed partial spectrum. Just as with the total spectrum, we can observe a very good agreement between theory and experiment and much sharper peaks and valleys in the CPE. The figure also shows that the CPE can describe the essential physics far away from the Fermi-surface rather well, as claimed in the ED section. This is not surprising, since the CPE produces a smooth self-energy on the real-axis, which is consistent with the basic assumption of a mean field theory such as the DMFT.

\section{Summary and Conclusions}\label{concsec}

We have presented a new algorithm, the continuous pole expansion (CPE), to analytically continue the self-energy of quantum many-body systems from (complex) Matsubara frequencies to the real axis. This method allows straightforward computation of electronic spectra for lattice models of strongly correlated systems from self-energy data collected in dynamical mean field (DMFT) simulations. The need for such an algorithm arises from developments of new, efficient continuous time quantum Monte Carole solvers for DMFT, which, in conjunction with non-equidistant Fast Fourier Transform, allow direct accumulation of the Greens function and self-energy on the Matsubara frequencies with controlled accuracy.

Since analytical continuation of complex functions is notoriously unstable, we have developed the CPE algorithm on the basis of two well-known analytic properties of the self-energy: (1) its imaginary part has a branch cut along the real axis and is negative definite in the upper complex plane (i.e. $\lim_{\varpi \rightarrow0}\rm{Im}[\Sigma(\vec{k}, \omega+\imath\,\varpi)] \leq 0$); and (2) it is analytic and has no poles in the upper complex plane. Thus, as a consequence of the first property the imaginary part of the self-energy can be parametrized as a purely negative function, which is a strong constraint, and the analyticity allows use of the Kramers-Kronig relationship to compute the self-energy everywhere in the complex plane from the imaginary part on the real axis. One can hence compute the self-energy on the Matsubara frequencies for any given parametrization of the imaginary part on the real axis. With this in mind, the CPE algorithm can be summarized in one sentence: {\it It consists of finding a negative definite parametrization of the imaginary part of the self-energy on the real frequency axis in such as way that the difference between accumulated QMC self-energy (data) and the computed self-energy on the Matsubara frequencies is minimized}.

With this the analytical continuation problem has been cast into a contained minimization problem. Minimization is much more stable numerically, especially with regard to statistical noise in the data that arises from the Monte Carlo sampling. Moreover, in the present case, the constrained minimization can be formulated as a quadratic programmable optimization with linear constraints. The latter is a well known problem for which many numerical algorithms exist. In the present paper we have used the simplest, the Frank-Wolf algorithm. 

Extensive validation of the CPE algorithm has been given in this paper, both in terms of exactly solvable finite size models as well as for lattice problems that are well known in the literature. The strengths and weaknesses of the CPE have been analyzed in terms of a direct comparison with exact solutions for an isolated 8-site cluster. The CPE reproduces very well the features of the spectral function near the Fermi energy. Farther away form the Fermi level, the CPE only reproduces broad features and is unable to track the sharp features of the spectral functions that result from individual eigenvalues of the Greens function matrix of the finite size system.

The first of two lattice problems used for the validation of the CPE algorithm is the 2D Hubbard model at half filling, and in particular the momentum-dependent gap formation for which many results based on the identity in Eq.~(\ref{cheating}) have been published recently. This identity has been validated with the CPE over a wide temperature range and our results agree with the literature. Furthermore, the CPE allowed us to conclude from the k-dependent spectrum that the gap is indeed smaller at the nodal than at the antinodal points. 

Finally, the CPE was used to compute the spectrum of $NiO$, a prototypical strongly correlated materials that has been extensively studied in experiment and simulation. The standard method of LDA+DMFT to compute the self-energy has been applied and the spectra computed with the CPE algorithm are in excellent agreement with XPS, XES, and BIS measurements on NiO published in the literature (see Fig.~\ref{fig:NiO_XPS}). This demonstrates that the CPE can be used as a robust, unambiguous method to compete spectral function of real materials from self-energy data that has been collected on the Matsubara frequencies in effective medium based quantum Monte Carlo simulations.

\bibliography{references}
\end{document}